%% file: main.tex
\documentclass[11pt, a4paper, twocolumn, goog]{google}

\usepackage[authoryear, sort&compress, round]{natbib}
\usepackage{multirow}
\usepackage{array}
\usepackage{tikz}
\usetikzlibrary{arrows.meta,positioning,shapes.geometric,calc,fit,backgrounds,decorations.pathreplacing}
\usepackage{xspace}

\newcommand{\system}{Muscle Memory\xspace}
\newcommand{\fullname}{Muscle Memory\xspace}


\title{Muscle Memory for Agents: Compile not Merely Retrieve}

\correspondingauthor{pgomran@google.com}

\reportnumber{} 

\renewcommand{\today}{2026}

\author[1]{Pouya Ghiasnezhad Omran}
\author[1]{Soujanya Lanka}
\author[1]{Qin Zhang}
\author[1]{Tanya Dixit}

\affil[1]{Google Cloud FDE}

\begin{abstract}
Memory for LLM agents has converged on a single architectural pattern: store experience as text, embeddings, reflections, or rules; retrieve at inference time; let a general-purpose orchestrator interpret what to do. This paper argues that the pattern is the wrong default for personalization. We position \fullname---the practice of compiling recurring user intent into purpose-built specialist agents---as a distinct memory paradigm from retrieval, and we argue that compilation is a better fit for the workloads where current assistants impose a \emph{multi-turn tax} on their users: making them repeatedly correct format, depth, and scope to obtain a domain-appropriate answer.

We support the position with a reference implementation and empirical evidence. The implementation is a four-phase pipeline (Harvest $\to$ Analyze $\to$ Augment $\to$ Evaluate) that mines conversational history, separates behavioral from task patterns, and emits quality-gated executable \emph{compiled specialists} with two-stage trigger matching. On 90 held-out scenarios across five user personas, the augmented assistant wins \textbf{32 of 36} cases where a specialist fires, an \textbf{88.9\% win rate}, with a \textbf{+2.05 personalization gain} and only a \textbf{$-$0.28 accuracy cost} on a 1--4 scale. We discuss why compilation is better suited than retrieval in this regime, what the result implies for the broader memory design space, and what open problems remain.
\end{abstract}

\begin{document}

\maketitle

\section{Introduction}
\label{sec:intro}

LLM-based assistants have proven to be a flexible and powerful
approach to a wide range of tasks, from code generation to business
planning. However, a fundamental limitation persists: they treat every
user identically and every conversation is bootstrapped. Consider two users of the same assistant. A senior
software engineer debugging a web service handler expects a terse,
code-heavy response with root cause, fix, and edge cases, all in one
turn. A cafe owner asking about quarterly cash-flow projections needs
bite-sized, reassuring guidance with no jargon. Both receive the same
verbose, generic output and spend three to four additional turns
correcting format, depth, and scope before reaching a useful answer for their domain-specific task.

We refer to this recurring cost as the \emph{multi-turn tax}. It is
not merely an inconvenience but a structural limitation of current
LLM assistants. Users repeat the same corrective instructions session
after session because the system never learns that this engineer
always wants defensive coding tips, or that this cafe owner gets
overwhelmed by long lists.

\paragraph{The position.}
The dominant response to this problem has been to give the orchestrating LLM more, or better, memory: raw transcripts via retrieval-augmented generation \citep[RAG;][]{Lewis2020}, virtual memory hierarchies \citep[MemGPT;][]{Packer2024}, verbal self-reflections \citep[Reflexion;][]{Shinn2023}, Zettelkasten-style reasoning memory \citep[A-Mem;][]{AMem2025}, or callable skill libraries \citep[Voyager;][]{Wang2023voyager} and agentic skills~\citep{SoKSkills2026}. However they differ in what they store, all share a single architectural pattern: \emph{store, retrieve, let the orchestrator interpret}. The orchestrator is asked to translate retrieved memories into action within its own context window and inference budget.

We argue this default is not enough for personalization. Conversational history should not just be retrieved; it should also be \emph{compiled}---into purpose-built specialist agents, each owning an entire recurring task class, each carrying its own LLM call, prompt, and (where useful) multi-step blueprint, and each tested against real user history before deployment. We call this paradigm \fullname. Where retrieval asks the orchestrator to do more work with more context, compilation removes the orchestrator from the critical path entirely: its role reduces to trigger matching and delegation.

This is analogous to how humans develop \emph{muscle memory}. A pianist learning a sonata initially recruits conscious cognitive control over every finger; with practice, the same passages are committed to motor programs that execute without deliberation. The orchestrator-with-memory paradigm leaves every inference in the conscious-cognition regime. Compiled Memory moves recurring patterns to the muscle-memory regime, where the costly, error-prone step of re-deriving the right behavior from raw context is replaced by a tested, fast, automatic reflex.

The core insight that makes this possible is that follow-up turns in recurring conversations reveal the user's abstract goal: the corrections, clarifications, and pivots are not noise but a signal of what the assistant should have delivered upfront. \system extracts these signals and compiles them into compiled specialists that activate at runtime to deliver complete, preference-aligned responses in one or two turns.

We argue that conversational history is an underutilized resource.
Rather than storing and retrieving raw fragments, AI systems should
extract behavioral patterns and compile them into executable agent
code that anticipates user needs proactively. We validate this position empirically on five diverse user personas.

\paragraph{Contributions.}
\begin{enumerate}
  \item A \textbf{position}: compiled memory should be treated as a first-class memory paradigm, distinct from retrieval-based approaches, and the right default for personalization workloads.
  \item A \textbf{reference implementation}: a four-phase pipeline (Harvest $\to$ Analyze $\to$ Augment $\to$ Evaluate) that transforms conversational history into a swarm of compiled specialists (available at\footnote{\url{https://github.com/GoogleCloudPlatform/generative-ai/tree/main/agents/personalized-agent-swarms}}).
  \item A set of \textbf{techniques} that, we argue, recur in any compiled-memory system: \emph{behavioral/task pattern separation} to prevent style cues from triggering task agents; \emph{two-stage trigger matching} combining embedding similarity with per-agent binary questionnaires; \emph{task-adaptive style dampening} that overrides format constraints for complex tasks; and a \emph{three-layer hallucination guard} for cold-start safety.
  \item \textbf{Empirical evidence} across 90 held-out scenarios on five user personas, achieving an 88.9\% win rate with a +2.05 personalization gain---offered as evidence for the position, not merely for the system.

\end{enumerate}

\section{Related Work / The Retrieval Paradigm and Its Limits}
\label{sec:related}

\subsection{Conversational Memory, Reasoning Memory, and Personalization}

Endowing LLMs with long-term memory has attracted growing attention
in recent years, with methods evolving from raw storage toward
increasingly structured memory representations.
MemGPT~\citep{Packer2024} introduces a virtual memory hierarchy
inspired by operating-system paging, enabling LLMs to manage context
beyond fixed windows.
LongMem~\citep{Wang2023longmem} augments language models with a
retrieval-based long-term memory that decouples storage from model
parameters.
At the application level, Persona-Chat~\citep{Zhang2018} grounds
dialogue in explicit persona descriptions, while
LaMP~\citep{Salemi2024} provides a benchmark for language model
personalization across diverse user-facing tasks, and
PEARL~\citep{Mysore2023} personalizes writing assistants via
generation-calibrated retrievers.

More recently, \emph{reasoning memory} systems have moved beyond
storing raw text toward capturing problem-solving strategies and
procedural knowledge. Reflexion~\citep{Shinn2023} stores verbal
self-reflections of task failures and re-injects them on subsequent
attempts, achieving reinforcement without weight updates.
A-Mem~\citep{AMem2025} uses Zettelkasten-inspired dynamic indexing to
create interconnected knowledge networks from experience.
\citet{Zhang2024survey} provide a comprehensive survey of agent
memory mechanisms, organizing them by source, form, and operation,
while \citet{Wu2025procedural} formalize agent skills as
procedural memory and propose a four-stage skill lifecycle
(acquisition, representation, invocation, and refinement).
LATS~\citep{Zhou2024lats} unifies reasoning, acting, and planning via
Monte Carlo Tree Search with LLM-powered value functions, storing
failed trajectories and self-reflections as additional context for
subsequent iterations on the same task.

Among the runtime memory systems discussed above, MemGPT, Reflexion,
A-Mem, PEARL, and LATS share a common architectural pattern: they
enrich the memory of a \emph{single orchestrating LLM}, which remains
responsible for interpreting retrieved memories, reasoning over them,
and generating the final output. Whether the memory contains raw text
(MemGPT), reflections (Reflexion), or linked experience notes (A-Mem),
the orchestrator must translate memory into action within its own
context window and inference budget. LongMem is a notable exception:
its trained side-network fuses cached key-values into the generation
process rather than injecting retrieved text into the orchestrator's
context window. \system departs from the retrieve-and-interpret
paradigm entirely. Rather than retrieving instructions or reasoning traces and feeding
them back to a general-purpose orchestrator, \system \emph{compiles}
recurring patterns into standalone executable mini-agents, each with
its own dedicated LLM call, model parameters, enriched prompt, and
optional agentic blueprint (e.g., critic/generative loops, multi-step
pipelines such as diagnose $\to$ recommend $\to$ format). The
orchestrator does not interpret the memory; it \emph{delegates} the
task to a purpose-built agent that has been tested against real
conversation history before deployment.

\subsection{Skill Libraries and Agentic Skills}

A parallel line of work stores reusable \emph{skills} rather than
memories. Voyager~\citep{Wang2023voyager} maintains an ever-growing
library of executable JavaScript functions for Minecraft, indexed by
embedding similarity and retrieved at task time, with a
self-verification gate that admits only programs confirmed to achieve
their task. Recent work systematizes agentic
skills~\citep{SoKSkills2026} as callable modules with applicability
conditions, execution policies, and termination criteria, cataloguing
patterns from executable code to self-evolving libraries. Separately,
SWE-agent~\citep{Yang2024sweagent} defines specialized agent-computer
interfaces for code editing and navigation---a fixed, hand-authored
command set rather than a retrieved skill library. Coding assistants
such as Claude Code and Cursor allow users to define reusable
``skills'' or ``rules'' that are injected into the orchestrator's
context when triggered.

These approaches represent an important step toward executable
procedural knowledge. However, even when skill bodies execute in
external runtimes (as in Voyager's JavaScript interpreter), the
orchestrator must \emph{select, compose, and invoke} every skill
within its own context window and inference budget. For
prompt-injected and natural-language skill patterns---such as the
rules in Claude Code and Cursor, and the corresponding patterns in
\citeauthor{SoKSkills2026}'s~(\citeyear{SoKSkills2026}) taxonomy---the
skill's quality is additionally bounded by the orchestrator's ability
to follow instructions faithfully, which is particularly challenging
for complex, multi-step tasks where format constraints, domain
expertise, and quality requirements interact.

Our approach is fundamentally different. Each mini-agent is a
\emph{self-contained execution unit} with an axis of freedom from context accumulated so far with the main orchestrator: it makes its own LLM call with
a task-specific enriched prompt, can implement multi-step agentic
patterns (e.g., a financial-planning agent that drafts $\to$
validates assumptions $\to$ formats for a non-technical audience),
and has been quality-gated through critic passes, fact-checking, and
mini-eval validation before deployment. The orchestrator's role
reduces to trigger matching and delegation; the complex,
quality-critical work is offloaded to a tested, purpose-built agent
rather than being handled by the same general-purpose LLM that
processes all other requests. This architectural distinction,
\emph{delegation to tested specialists} versus \emph{instruction
retrieval for a generalist}, is what enables \system to achieve an
88.9\% win rate on personalized tasks without sacrificing accuracy.

\subsection{Multi-Agent Systems}

Recent frameworks orchestrate multiple LLM agents for complex task
solving. AutoGen~\citep{Wu2023} enables flexible multi-agent
conversations with customizable interaction patterns.
MetaGPT~\citep{Hong2024} assigns software-engineering roles
(architect, engineer, QA) to agents collaborating via structured
standard operating procedures. CAMEL~\citep{Li2023camel} explores
emergent behaviors through communicative agent role-playing, while
AgentVerse~\citep{Chen2023agentverse} investigates dynamic group
composition, automatically recruiting agents for each task.
Parallel to structured collaboration, \citet{Li2024moreagents} explore
scaling via sampling and voting with undifferentiated agent ensembles.
In frameworks such as AutoGen, MetaGPT, and CAMEL, agent roles are
defined per task---by a human designer or, in AgentVerse's case,
generated on the fly for the task at hand---rather than mined from a
specific user's interaction history. \system generates agents
\emph{automatically} from user data, with quality gating that
determines the optimal number and scope of agents per user (1--6 in
practice). Moreover, while these multi-agent frameworks focus on
\emph{decomposing a single complex task} across collaborating agents,
\system creates agents that each \emph{own an entire task class}, that
is, a recurring user need learned from history, and operate
autonomously without inter-agent coordination.

\subsection{LLM-as-Judge Evaluation}

Automated evaluation of LLM outputs increasingly relies on LLMs
themselves as judges. MT-Bench and Chatbot Arena~\citep{Zheng2023}
establish pairwise comparison as a scalable evaluation paradigm,
while AlpacaEval~\citep{Dubois2024} introduces length-controlled
automatic evaluation to reduce verbosity bias.
\citet{Wang2024faireval} identify position bias in LLM judges and
propose debiasing strategies. However, existing pairwise protocols compare responses on fixed prompt
sets without conditioning on a persistent per-user profile or letting
a persona-conditioned simulated user drive the conversation. Building
on work in simulated-user personalization
evaluation~\citep{Zhao2025personalens}, \system applies a
\emph{user-agent-driven} evaluation protocol where a simulated user
autonomously drives multi-turn conversations with both baseline and
augmented assistants under evaluation parity, assessing
personalization quality over extended interactions.

\section{The Compiled Memory Position}
\label{sec:position}

We define \fullname as the practice of converting recurring patterns from a user's interaction history into standalone executable specialist agents---each owning an entire task class, each invoked at runtime via trigger matching rather than prompt augmentation, and each tested before deployment. This section articulates the position as three principles, contrasts compiled memory with retrieved memory across the dimensions where they diverge, and surfaces the assumptions that determine when compilation is the right choice.

\subsection{Three principles}

\paragraph{P1: Compilation over retrieval.}
A pattern that recurs across sessions is evidence of a stable workload. Stable workloads should be compiled into code, not re-derived from retrieved fragments at every inference. Retrieval is appropriate for one-off facts whose interpretation depends on the live request; compilation is appropriate for recurring procedures whose interpretation is itself the part that should be cached.

\paragraph{P2: Specialists over generalists.}
A purpose-built agent with a task-specific enriched prompt, dedicated model parameters, and (optionally) a multi-step blueprint outperforms a generalist orchestrator that has been handed equivalent instructions and asked to follow them. The specialist is bounded by what it has been tested on, not by what an instruction-following generalist happens to attend to in its context window. A specialist also has an axis of freedom from the context accumulated so far with the main orchestrator: it begins each invocation with its own prompt, not the orchestrator's drift.

\paragraph{P3: Tested before deployment.}
A compiled specialist is an artifact: it can be quality-gated through critic passes, fact-checks, and held-out evaluations \emph{before} it is allowed to run in production. Some skill-based systems gate admission similarly---Voyager verifies task completion in a simulator with programmatic success signals---but retrieved-then-interpreted memory in open-ended conversational settings cannot be pre-tested in this way; its behavior emerges only at inference, conditioned on the orchestrator and the rest of the prompt, and is therefore difficult to audit, regression-test, or version.

\subsection{Compiled vs.\ retrieved: where they diverge}

Compiled and retrieved memory differ along several dimensions that matter for personalization workloads. Where retrieved memory ships textual context to the orchestrator at every call---whether raw passages~\citep{Lewis2020}, virtual memory pages~\citep{Packer2024}, or verbal reflections~\citep{Shinn2023}---compiled memory ships an executable artifact that is invoked instead of the orchestrator. Where retrieval quality is bounded by the orchestrator's instruction-following, compiled quality is bounded by the specialist's own prompt and blueprint, both of which are tested. Where retrieved-memory failures appear at inference and are hard to localize, compiled-memory failures show up as failed quality gates before deployment, or as miss-routes at trigger matching, both of which are debuggable in isolation.

The cost structure also differs. Both paradigms incur per-call overhead, but its nature and scaling differ. Retrieval embeds the query, searches a store, and injects retrieved fragments into the orchestrator's prompt, inflating input tokens on every call by an amount that grows with the breadth of stored experience. Compilation pays a lightweight, fixed-size routing cost (one feature-extraction call plus one embedding lookup; see Section~\ref{sec:augment}) that is independent of how many patterns have been compiled. Compilation additionally pays an up-front compile cost amortized over the lifetime of each specialist. The trade-off favors compilation precisely when patterns are stable and recurrence is high, the regime that personalization occupies.

\subsection{When compilation is the right choice}

Compiled memory is not a universal replacement for retrieval. It is the right default when:
\begin{itemize}
  \item The same intent recurs frequently enough across sessions to amortize compilation cost.
  \item Quality, style, or workflow consistency matters across instances of that intent.
  \item Patterns can be discovered from observed history rather than only being declared by the user up front.
\end{itemize}
Where these conditions fail---one-shot factual queries, novel tasks with no history, exploratory dialogue---retrieval (or no memory at all) remains the correct choice. The two paradigms are complementary, but for conversational personalization the field has defaulted to retrieval even where compilation would dominate.

\section{A Reference Implementation}
\label{sec:method}

\system operates as a sequential four-phase pipeline
(Figure~\ref{fig:architecture}). Each phase produces artifacts
consumed by the next: raw sessions $\to$ pattern-derived swarms $\to$
augmented runtime $\to$ comparative evaluation logs.

\input{architecture}

\subsection{Phase 1: History Harvest}
\label{sec:harvest}

We generate 250 synthetic multi-turn conversations across five user
personas spanning diverse domains: a senior software engineer, a
marketing manager, an ML graduate student, a small-business (cafe)
owner, and a travel/cooking enthusiast. Each persona has 50
predefined scenarios with a specific intent and one of four follow-up
strategies (\emph{clarify}, \emph{deep\_dive}, \emph{pivot},
\emph{correct}). A dual-LLM orchestration drives each session: a
user agent role-playing the persona (temperature 0.9 for diversity)
interacts with a standard assistant, producing naturalistic
multi-turn dialogues that capture realistic preference-expression
patterns.

\subsection{Phase 2: Pattern Analysis and Swarm Generation}
\label{sec:analyze}

The analysis pipeline extracts, validates, and compiles mini-agents
through eight stages.

\smallskip\noindent\textbf{Pattern extraction and classification.}\enspace
Sessions are batched (10 at a time) and analyzed by an LLM to
identify recurring patterns (frequency $\geq 3$), which are merged
via Jaccard similarity ($> 0.5$) to consolidate overlapping intents.
Patterns are classified as either \emph{task patterns} (what the user
wants: financial planning, code debugging) or \emph{behavioral
patterns} (how the user communicates: brevity preference, overwhelm
avoidance).

This separation is a key design decision. Treating behavioral
patterns identically to task patterns produces agents that trigger on
communication style cues rather than actual task intent, leading to
high false-positive rates. By separating behavioral patterns into a
shared style profile (\texttt{user\_style.json}) applied to all agents
at runtime, these false positives are eliminated entirely.

\smallskip\noindent\textbf{Agent generation and quality gating.}\enspace
Task patterns are compiled into executable Python modules with
enriched prompts containing baked-in user preferences. Agents follow
two architectures: \emph{static} (single LLM call with enriched
prompt) and \emph{dynamic} (multi-step pipeline, typically 2--3
stages such as diagnose $\to$ recommend $\to$ format).

To ensure quality, each agent passes through a multi-stage gate:
\begin{enumerate}[label=(\alph*)]
  \item \textbf{Parallel generation}: two candidates generated at
    temperatures 0.2 and 0.35; a lightweight LLM selects the more
    grounded version.
  \item \textbf{Post-generation fact-check}: scans the enriched
    prompt for fabricated claims; low-confidence assertions are
    revised with uncertainty caveats.
  \item \textbf{Critic pass}: validates each agent against real
    conversation history on 10 criteria with a 7/10 threshold;
    a fabrication hard-ceiling caps the score at 4/10 if $\geq 2$
    contradictions are found.
  \item \textbf{Non-parametric ranking}: scores agents on five
    weighted dimensions (value $\times 3$, distinctiveness $\times 2$,
    trigger clarity $\times 2$, quality $\times 2$,
    frequency $\times 1$); agents scoring $\geq 25/50$ are retained.
  \item \textbf{Overlap merge}: pairwise overlap detection
    (embedding cosine + domain/task-type Jaccard $\geq 0.73$) triggers
    an LLM merge-or-keep decision, preventing cross-domain misrouting
    among semantically similar agents.
  \item \textbf{Mini-eval validation}: a subset of the Phase~4
    evaluation harness tests each surviving agent on 6 historical
    scenarios (3 similar, 3 different); agents with average accuracy
    $< 2.5$ or zero trigger rate are pruned.
\end{enumerate}

\smallskip\noindent\textbf{Trigger pre-computation.}\enspace
For each agent, the pipeline pre-computes a 768-dimensional scope
embedding (text-embedding-005) and generates 3--5 contrastive binary
(yes/no) questionnaire items for runtime disambiguation.

\subsection{Phase 3: Runtime Augmentation}
\label{sec:augment}

At inference time, each user message passes through the active memory
layer (Figure~\ref{fig:runtime_eval}a).

\smallskip\noindent\textbf{Stage 1: Embedding and soft attribute filter.}\enspace
A single LLM call extracts structured features (domain, task type,
specificity, keywords, action-object). The user message is embedded
(768-dim) and scored against all agent scope embeddings via cosine
similarity, with soft penalties for attribute mismatches ($-0.15$ for
domain, $-0.10$ for task type). All agents above a threshold ($\geq
0.45$) pass to Stage~2.

Soft penalties, as opposed to hard binary rejection (exact domain
match required), allow cross-domain requests to match relevant agents
with reduced but non-zero scores, improving recall without sacrificing
precision.

\smallskip\noindent\textbf{Stage 2: Binary questionnaire disambiguation.}\enspace
When a single agent passes Stage~1, it is selected directly. When
multiple candidates survive, each candidate's pre-computed
questionnaire (3--5 yes/no questions) is evaluated against the user
message in parallel LLM calls. The combined score is 40\% embedding +
60\% questionnaire match ratio. If one agent achieves $\geq 0.8$
match ratio, it is selected. If multiple agents exceed this threshold,
the highest combined score wins. If none reach 0.8, a stronger LLM
tiebreaker is invoked.

The questionnaire stage is essential for domains with semantically
similar agents. For the cafe owner persona, embedding similarity
alone could not distinguish five cafe-domain agents (menu pricing
vs.\ supplier management vs.\ operations); the questionnaire stage
resolved all misroutes.

\smallskip\noindent\textbf{Routing overhead.}\enspace
  The per-call routing cost of the two-stage matching pipeline is bounded and predictable. Stage~1 requires exactly one lightweight LLM call for feature extraction (structured JSON output, ${\sim}500$ tokens) and one embedding API call, both fixed costs independent of swarm size. Stage~2 fires only when multiple agents survive Stage~1; when it does, questionnaire evaluations run in parallel, so wall-clock latency equals that of a single call regardless of the number of candidates. In contrast, a retrieval-based system embeds the query (comparable cost), performs a vector search, and injects retrieved context into the orchestrator's prompt, typically adding 500--2{,}000 tokens of input to the
  main LLM call on every invocation. Compilation's routing overhead replaces retrieval's
  context-injection overhead: a fixed-cost routing stage versus a variable-cost prompt inflation. For users with well-separated agent domains (e.g., user\_2, where Stage~2 rarely fires), the routing cost reduces to a single Flash call plus an embedding lookup.

  \smallskip\noindent\textbf{Task-adaptive style dampening.}\enspace
When the matched agent detects a complex task (financial plans,
pricing strategies, technical architecture), it automatically
overrides conflicting style constraints. For example, a user whose
style profile specifies ``bite-sized, one-concept-at-a-time''
responses would have a comprehensive financial plan fragmented across
multiple turns; style dampening preserves tone and language level
while allowing complete delivery.

\smallskip\noindent\textbf{Hallucination guard.}\enspace
Fabrication defense operates at two stages.
\emph{Generation-time:} the critic pass scores fabricated references
and technical claims as critical failures (score~0); agents that
exhibit fabrication in the validation gate are pruned before
deployment.
\emph{Runtime:} on first-turn messages only (where session history is
naturally empty), regex scanning detects fabricated-context markers
(e.g.\ ``as we discussed,'' ``your message was cut off'') and falls
back to the baseline response. In multi-turn conversations,
mini-agents receive the full session history and these checks are
bypassed, avoiding false suppression of valid personalized references.

\subsection{Phase 4: Evaluation Protocol}
\label{sec:eval}

We employ a user-agent-driven evaluation protocol
(Figure~\ref{fig:runtime_eval}b). For each of 90 held-out scenarios (18
per user: 10 similar to training intents, 8 different), a simulated
user agent sends the same opening message to both the baseline and
augmented assistants and autonomously drives multi-turn conversations
until the goal is reached or a turn cap is hit.

Both agents run under \emph{fair evaluation parity}: neither has
access to web search or external memory tools; the only difference is
the augmented agent's swarm tool. An LLM judge (Gemini 3.1 Pro)
scores both transcripts on three dimensions, each on a 1--4 scale:
\emph{accuracy} (factual correctness; fabricated technical claims
cap the score at~1), \emph{helpfulness} (goal completion), and
\emph{personalization} (alignment with learned preferences).
The judge then declares a winner.

\begin{figure*}[t]
\centering
\begin{minipage}[t]{0.48\textwidth}
\centering
\input{runtime}
\centerline{\small (a)}
\end{minipage}\hfill
\begin{minipage}[t]{0.48\textwidth}
\centering
\input{evaluation}
\centerline{\small (b)}
\end{minipage}
\caption{%
\textbf{(a)} Runtime message flow through the active memory layer with two-stage matching.
\textbf{(b)} Evaluation protocol: same opening message drives parallel conversations scored by an LLM judge.
}
\label{fig:runtime_eval}
\end{figure*}

\section{Empirical Evidence}
\label{sec:experiments}

\subsection{Setup}

The complete source code for our pipeline is publicly available\footnote{\url{https://github.com/GoogleCloudPlatform/generative-ai/tree/main/agents/personalized-agent-swarms}}. Table~\ref{tab:users} summarizes the five user personas, their
domains, and the number of mini-agents generated by the quality-gated
pipeline. The total of 23~agents across 5~users reflects the
non-parametric selection: no fixed target is imposed, and only agents
scoring $\geq 25/50$ on the ranking rubric survive. Evaluation uses
90 held-out scenarios (50~similar, 40~different) with an LLM judge
(Gemini 3.1 Pro, temperature~1.0).

\begin{table}[t]
\centering
\caption{User personas and generated mini-agent counts.}
\label{tab:users}
\small
\resizebox{\columnwidth}{!}{%
\begin{tabular}{@{}llcl@{}}
\toprule
\textbf{User} & \textbf{Persona} & \textbf{Agents} & \textbf{Domain} \\
\midrule
user\_1 & Software engineer & 1 & Python, GCP, Docker \\
user\_2 & Marketing manager & 6 & Copy, metrics, strategy \\
user\_3 & ML grad student   & 6 & NLP, training, math \\
user\_4 & Cafe owner        & 5 & Finance, ops, suppliers \\
user\_5 & Travel + cooking  & 5 & Trips, recipes, culture \\
\midrule
\multicolumn{2}{@{}l}{\textbf{Total}} & \textbf{23} & \\
\bottomrule
\end{tabular}%
}
\end{table}

Our experiments are designed to validate the following observations:
\begin{enumerate}
  \item \system significantly outperforms the baseline assistant in
    terms of personalization when mini-agents fire.
  \item The accuracy cost of personalization is minimal, indicating
    that the quality-gating pipeline produces reliable agents.
  \item The pipeline components (behavioral separation, embeddings,
    hallucination guard, questionnaires) are complementary; no single
    component subsumes the others.
\end{enumerate}

\subsection{Main Results}

Table~\ref{tab:results} presents head-to-head results when
mini-agents fire. Across all users, the augmented assistant wins
\textbf{32 of 36} scenarios where a swarm agent activates, achieving
an \textbf{88.9\% win rate}. Personalization improves by
\textbf{+2.05} on a 1--4 scale (from 1.67 to 3.72) with a modest
accuracy cost of \textbf{$-$0.28} (from 3.92 to 3.64).

\begin{table}[t]
\centering
\caption{\textbf{Head-to-head results when mini-agents fire.} Accuracy and personalization scores on 1--4 scale for similar-domain scenarios. $B \rightarrow A$ = baseline $\rightarrow$ augmented. \textit{(Excludes 8 false-positive firings).}}
\label{tab:results}
\small
\resizebox{\columnwidth}{!}{%
\begin{tabular}{@{}lcccc@{}}
\toprule
\textbf{User} & \textbf{Fired} & \textbf{W-L-T} &
  \textbf{Acc.\ $\Delta$} & \textbf{Pers.\ $\Delta$} \\
\midrule
user\_1 &  5/10 & 3-2-0 & $-1.00$ & $+1.20$ \\
user\_2 & 10/10 & \textbf{10-0-0} & $\phantom{-}0.00$ & $+2.50$ \\
user\_3 &  9/10 & 8-1-0 & $-0.56$ & $+2.23$ \\
user\_4 &  6/10 & \textbf{6-0-0} & $\phantom{-}0.00$ & $+2.17$ \\
user\_5 &  6/10 & 5-1-0 & $\phantom{-}0.00$ & $+1.67$ \\
\midrule
\textbf{All} & \textbf{36/50} & \textbf{32-4-0} &
  $\boldsymbol{-0.28}$ & $\boldsymbol{+2.05}$ \\
\bottomrule
\end{tabular}%
}
\end{table}

Two users are particularly noteworthy. \textbf{user\_2} (marketing
manager) achieves a perfect 10--0 record with zero accuracy loss and
$+2.50$ personalization, the highest individual gain. This user's six
agents cover well-separated task domains (A/B testing, metrics
analysis, email campaigns, social media, presentations, strategy),
making trigger matching unambiguous. \textbf{user\_4} (cafe owner)
achieves a 6W--0L record despite having five semantically similar
cafe-domain agents; the binary questionnaire disambiguation stage
successfully distinguishes these overlapping agents.

Trigger accuracy across all users is \textbf{72\%} (36/50 similar
scenarios fire). The false-positive rate is \textbf{20\%} (8/40
different-domain scenarios trigger an agent). However, these false
positives are rarely damaging: the augmented assistant still wins 5
of 8 false-positive cases.

\smallskip\noindent\textbf{Design lessons.}\enspace
The following observations are qualitative lessons from iterative development, not formal ablations; we report them because each failure mode motivated a specific architectural component. The current pipeline is the result of iterative, failure-driven refinement. Early designs that relied on keyword matching for agent
triggering produced poor precision. Treating behavioral patterns as
task patterns inflated false-positive rates. Hard binary domain
filters prevented legitimate cross-domain matches. Omitting
hallucination guards allowed fabricated first-turn context. Each of
these failure modes motivated a specific component in the final
design: scope embeddings, behavioral/task separation, soft attribute
penalties, and the two-stage hallucination defense (generation-time
pruning plus runtime guard), respectively.

\section{Discussion and Limitations}
\label{sec:discussion}

\noindent\textbf{Where augmentation excels.}\enspace
The strongest results emerge for users with well-separated task
domains (user\_2: marketing, 6 agents, 10--0) and users who express
strong format preferences that a generic assistant consistently
violates (user\_4: non-technical, needs reassurance, 6--0). The
hybrid paradigm, standard conversation for most interactions with
seamless mini-agent activation for recurring patterns, reduces
multi-turn friction without sacrificing flexibility.

\smallskip\noindent\textbf{Where it struggles.}\enspace
user\_1 (software engineer) was over-pruned to a single agent,
limiting coverage. The non-parametric ranking penalized agents whose
scope overlapped with general-purpose LLM capabilities (``explain
Python errors'' is valuable to the user but indistinct from the
baseline). The 20\% false-positive rate on different-domain scenarios,
while rarely damaging, indicates that embedding similarity alone is
insufficient for hard negatives; future work could introduce explicit
rejection questionnaires.

\smallskip\noindent\textbf{Runtime guard limitations.}\enspace
The runtime regex layer covers common fabricated-context phrases but
is not exhaustive against diverse LLM phrasings. We chose
deterministic regex for zero-latency inference; false-positive risk is
bounded by restricting checks to first-turn messages only. A
lightweight LLM-based classifier as a complementary layer is a
promising direction.

\smallskip\noindent\textbf{Limitations.}\enspace
We acknowledge several limitations. First, all evaluation is
synthetic: simulated user agents and LLM judges, not human
participants. While the user-agent-driven protocol and fair
evaluation parity provide more rigorous synthetic evaluation than
static benchmarks---prior work on simulator validity~\citep{Dou2025simulatorarena} reports Spearman correlations of up to 0.7 with human judgments on multi-turn tasks---a human evaluation study is essential future work. Second, the pipeline uses a single model family (Gemini) for
generation, matching, and evaluation; cross-model validation would
strengthen generalizability claims. Third, agent triggers are static
after generation, that is, the system does not adapt agents based on
runtime feedback or evolving user preferences. Fourth, there is no
cross-user pattern transfer; each user's swarm is generated
independently despite potential shared patterns across users with
similar roles.

\smallskip\noindent\textbf{Accuracy-personalization trade-off.}\enspace
The aggregate accuracy cost ($-0.28$ for $+2.05$) is concentrated:
user\_1 incurs $-1.00$ (single agent, limited coverage), user\_3
incurs $-0.56$ (specialized ML API surface), while user\_2, user\_4,
and user\_5 show zero cost---suggesting the deficit correlates with
domain technicality rather than architectural isolation.

Error analysis reveals a consistent pattern: mini-agents produce
stylistically personalized responses but hallucinate \emph{technical}
details. For instance, a statistics agent wrote
\texttt{mannwhitneyu(\ldots, continuity=True)} instead of the correct
\texttt{use\_continuity}, crashing at runtime (baseline: 4/4,
augmented: 1/4). This is distinct from the conversational fabrication
the runtime guard targets (``as we discussed''); the generation-time
critic catches many such cases, but specialized API surfaces remain
challenging. Retrieval-augmented verification of technical claims and
user-controllable personalization levels are promising mitigations. We note that the current quality gates target general factual grounding — verifying that claims are not fabricated and that outputs align with user history — but do not validate domain-specific API correctness. Closing this gap likely requires tool-augmented verification (e.g., executing generated code snippets in a sandbox) rather than LLM-only critic passes.  

\smallskip\noindent\textbf{Runtime cost trade-offs.}\enspace
  The routing pipeline introduces per-call overhead that we do not claim is zero. Every query pays one Flash feature-extraction call and one embedding lookup. When multiple agents compete, parallel questionnaire calls add wall-clock latency equivalent to a single Flash call. The Pro tiebreaker, reserved for cases where no questionnaire achieves a ${\geq}0.8$ match ratio, fires infrequently. Importantly, this routing overhead is \emph{architecturally bounded}: it does not grow with the number of compiled agents or the depth of stored history. This contrasts with retrieval-based approaches, where per-call cost grows with the amount of context injected, a property that becomes increasingly costly as memory accumulates over long user histories. A full empirical latency and token-cost comparison between compilation routing and retrieval-based alternatives is a valuable direction for future work; we note that such a comparison must account for retrieval's own per-call costs (embedding, search, context injection), which are often underreported in retrieval-system analyses.

\smallskip\noindent\textbf{Scope of the position.}\enspace
We do not claim that compiled memory subsumes retrieval. We claim that it is currently under-explored relative to its fit for personalization workloads---concurrent work on agent-skill compilation for cross-harness efficiency~\citep{Chen2026skvm} targets a different regime---and that for conversational personalization the field has defaulted to retrieval in regimes where compilation would dominate. The right system is plausibly a hybrid: compiled specialists for recurring intents, retrieval for one-off facts, and a router that knows the difference.

\section{Conclusion and Future Work / Implications and Open Problems}
\label{sec:conclusion}

In this paper, we have presented \fullname, a pipeline that extracts
personalized mini-agent swarms from conversational history and
demonstrated that this approach achieves an 88.9\% win rate with a
+2.05 personalization gain at minimal accuracy cost. Our central
argument, that conversational history should produce executable
agents rather than merely retrievable memories, is supported by the
experimental results across 90 held-out scenarios, where the full
pipeline achieves a 32W--4L record.

The behavioral/task separation principle, two-stage trigger matching,
and task-adaptive style dampening are general-purpose techniques
applicable beyond our specific pipeline. We believe they represent
foundational building blocks for personalized multi-agent systems.

There are several interesting directions for future work. We plan to
conduct a human evaluation study replacing simulated users and LLM
judges with real participants. We also plan to investigate continuous
learning, where agents adapt based on runtime feedback rather than
remaining static after generation. Cross-user pattern transfer,
enabling shared agents for users with similar professional roles and adaptive personalized harness, is
another promising direction. Finally, model-agnostic validation
across multiple LLM families and per-user adaptive thresholds for
the accuracy-personalization trade-off based on domain sensitivity
remain open problems.

\section*{Acknowledgments}
We would like to thank Ashmita Kapoor for her valuable feedback and for reviewing this paper.

\section*{Declaration on Generative AI}
During the preparation of this work, the author(s) used LLM-based
tools in order to: assist with drafting, perform grammar and spell checks, assist with LaTeX formatting and bibliography
management. After using these tools, the author(s) reviewed and
edited the content as needed and take full responsibility for the
publication's content.

\bibliography{main}

\end{document}

%% file: architecture.tex

\begin{figure*}[t]
\centering
\resizebox{\textwidth}{!}{%
\begin{tikzpicture}[
  >=Stealth,
  font=\sffamily,
  phase/.style={
    rounded corners=8pt, draw=#1, line width=1.5pt,
    fill=#1!6, minimum width=2.6cm, minimum height=1.1cm,
    align=center, font=\sffamily\bfseries\footnotesize
  },
  substep/.style={
    rounded corners=5pt, draw=#1, line width=0.8pt,
    fill=#1!8, minimum width=2.2cm, minimum height=0.6cm,
    align=center, font=\sffamily\scriptsize
  },
  phasearrow/.style={->, line width=2pt, color=#1},
  subarrow/.style={->, line width=0.7pt, color=gray!60},
  phaselabel/.style={font=\sffamily\bfseries\small, color=#1},
  annot/.style={font=\sffamily\tiny, color=gray!70},
  dashbox/.style={rounded corners=6pt, draw=gray!40, dashed, line width=0.8pt, inner sep=8pt},
]

\definecolor{cHarvest}{HTML}{0D9488}
\definecolor{cAnalyze}{HTML}{7C3AED}
\definecolor{cAugment}{HTML}{2563EB}
\definecolor{cEval}{HTML}{DC2626}
\definecolor{cGreen}{HTML}{059669}
\definecolor{cAmber}{HTML}{D97706}

\node[phase=cHarvest] (harvest) at (0,0) {HARVEST};
\node[phaselabel=cHarvest, above=2pt of harvest] {Phase 1};
\node[annot, below=2pt of harvest] {5 users $\times$ 50 sessions};

\node[phase=cAnalyze] (analyze) at (6.5,0) {ANALYZE};
\node[phaselabel=cAnalyze, above=2pt of analyze] {Phase 2};

\node[phase=cAugment] (augment) at (13,0) {AUGMENT};
\node[phaselabel=cAugment, above=2pt of augment] {Phase 3};
\node[annot, below=2pt of augment] {Runtime};

\node[phase=cEval] (eval) at (19,0) {EVALUATE};
\node[phaselabel=cEval, above=2pt of eval] {Phase 4};
\node[annot, below=2pt of eval] {90 scenarios};

\draw[phasearrow=cAnalyze] (harvest.east) -- (analyze.west);
\draw[phasearrow=cAugment] (analyze.east) -- (augment.west);
\draw[phasearrow=cEval] (augment.east) -- (eval.west);

\begin{scope}[shift={(0,-2.4)}]
  \node[substep=cHarvest] (h1) at (0,0) {User Agent};
  \node[annot, below=1pt of h1] {\itshape 5 personas};

  \node[substep=cHarvest] (h2) at (0,-1.1) {Assistant Agent};
  \node[annot, below=1pt of h2] {\itshape generic LLM};

  \node[substep=cHarvest] (h3) at (0,-2.2) {Session Logs};
  \node[annot, below=1pt of h3] {\itshape 250 total};

  \draw[subarrow] (h1.south) -- (h2.north);
  \draw[subarrow] (h2.south) -- (h3.north);

  \begin{scope}[on background layer]
    \node[dashbox, fit=(h1)(h2)(h3), inner sep=12pt, fill=cHarvest!3] (harvestbox) {};
  \end{scope}
\end{scope}

\draw[subarrow, dashed] (harvest.south) -- ++(0,-0.6) -| (harvestbox.north);

\begin{scope}[shift={(5.2,-2.4)}]
  \node[substep=cAnalyze] (ae1) at (0,0) {Pattern Extraction};
  \node[annot, right=3pt of ae1.east, anchor=west] {Jaccard $>$ 0.5};

  \node[substep=cAmber] (ae2) at (0,-1.1) {Task / Behavioral Split};

  \node[substep=cAugment] (ae3a) at (-0.7,-2.2) {Agent Generation};
  \node[annot, below=1pt of ae3a] {\itshape static \& dynamic};
  \node[substep=cGreen] (ae3b) at (1.8,-2.2) {Style Profile};
  \node[annot, below=1pt of ae3b] {\itshape user\_style.json};

  \node[substep=cEval] (ae4) at (-0.7,-3.4) {Quality Gates};
  \node[annot, below=1pt of ae4, text width=2.5cm, align=center] {\itshape fact-check $\cdot$ critic $\cdot$ ranking};

  \node[substep=cGreen] (ae5) at (-0.7,-4.6) {Validation Gate};
  \node[annot, below=1pt of ae5] {\itshape mini-eval harness};

  \node[substep=cAnalyze] (ae6) at (-0.7,-5.7) {Final Swarm};
  \node[annot, below=1pt of ae6] {\itshape 1--6 agents/user};

  \draw[subarrow] (ae1.south) -- (ae2.north);
  \draw[subarrow] (ae2.south) -- ++(0,-0.25) -| (ae3a.north);
  \draw[subarrow] (ae2.south) -- ++(0,-0.25) -| (ae3b.north);
  \draw[subarrow] (ae3a.south) -- ++(0,-0.35) -- (ae4.north);
  \draw[subarrow] (ae4.south) -- ++(0,-0.15) -- (ae5.north);
  \draw[subarrow] (ae5.south) -- ++(0,-0.15) -- (ae6.north);

  \begin{scope}[on background layer]
    \node[dashbox, fit=(ae1)(ae2)(ae3a)(ae3b)(ae4)(ae5)(ae6), inner sep=12pt, fill=cAnalyze!3] (analyzebox) {};
  \end{scope}
\end{scope}

\draw[subarrow, dashed] (analyze.south) -- ++(0,-0.6) -| (analyzebox.north);

\begin{scope}[shift={(11.8,-2.4)}]
  \node[substep=cAugment] (ag1) at (0,0) {Feature Extraction};
  \node[annot, right=3pt of ag1.east, anchor=west] {1 LLM call};

  \node[substep=cAmber] (ag2) at (0,-1.1) {Stage 1: Embedding};
  \node[annot, right=3pt of ag2.east, anchor=west] {cosine + soft penalties};

  \node[substep=cAnalyze] (ag3) at (0,-2.2) {Stage 2: Questionnaire};
  \node[annot, right=3pt of ag3.east, anchor=west] {3--5 yes/no per agent};

  \node[substep=cGreen] (ag4) at (0,-3.3) {Agent Execute};
  \node[annot, right=3pt of ag4.east, anchor=west] {+ style dampening};

  \node[substep=cEval] (ag5) at (0,-4.3) {Hallucination Guard};
  \node[annot, right=3pt of ag5.east, anchor=west] {3-layer defense};

  \draw[subarrow] (ag1.south) -- (ag2.north);
  \draw[subarrow] (ag2.south) -- (ag3.north);
  \draw[subarrow] (ag3.south) -- (ag4.north);
  \draw[subarrow] (ag4.south) -- (ag5.north);

  \begin{scope}[on background layer]
    \node[dashbox, fit=(ag1)(ag2)(ag3)(ag4)(ag5), inner sep=12pt, fill=cAugment!3] (augmentbox) {};
  \end{scope}
\end{scope}

\draw[subarrow, dashed] (augment.south) -- ++(0,-0.6) -| (augmentbox.north);

\begin{scope}[shift={(19,-2.4)}]
  \node[substep=cEval] (ev1) at (0,0) {Same Opening};
  \node[substep=gray] (ev2a) at (-1,-1.1) {\textcolor{gray}{Baseline}};
  \node[substep=cAugment] (ev2b) at (1,-1.1) {\textcolor{cAugment}{Augmented}};
  \node[substep=cEval] (ev3) at (0,-2.2) {LLM Judge};
  \node[annot, below=1pt of ev3] {\itshape 3 dims $\times$ 1--4 scale};

  \draw[subarrow] (ev1.south) -- ++(0,-0.2) -| (ev2a.north);
  \draw[subarrow] (ev1.south) -- ++(0,-0.2) -| (ev2b.north);
  \draw[subarrow] (ev2a.south) -- ++(0,-0.2) -| (ev3.north);
  \draw[subarrow] (ev2b.south) -- ++(0,-0.2) -| (ev3.north);

  \begin{scope}[on background layer]
    \node[dashbox, fit=(ev1)(ev2a)(ev2b)(ev3), inner sep=12pt, fill=cEval!3] (evalbox) {};
  \end{scope}
\end{scope}

\draw[subarrow, dashed] (eval.south) -- ++(0,-0.6) -| (evalbox.north);

\node[annot, above=0pt] at ($(harvest.east)!0.5!(analyze.west)+(0,0.15)$) {sessions};
\node[annot, above=0pt] at ($(analyze.east)!0.5!(augment.west)+(0,0.15)$) {swarm};
\node[annot, above=0pt] at ($(augment.east)!0.5!(eval.west)+(0,0.15)$) {logs};

\end{tikzpicture}
}%
\caption{
\textbf{Muscle Memory system architecture.}
Four sequential phases produce personalized mini-agent swarms from conversational history.
\textbf{Harvest} generates 250 synthetic multi-turn sessions across 5 user personas.
\textbf{Analyze} extracts recurring patterns, separates task from behavioral signals,
generates executable agents with multi-stage quality gating (fact-check, critic pass, non-parametric ranking, merge, mini-eval validation), and pre-computes scope embeddings and binary questionnaires.
\textbf{Augment} deploys agents at runtime via two-stage matching
(embedding similarity + questionnaire disambiguation) with hallucination guard and task-adaptive style dampening.
\textbf{Evaluate} runs 90 held-out head-to-head comparisons with an LLM judge.
}
\label{fig:architecture}
\end{figure*}

%% file: runtime.tex

\resizebox{\linewidth}{!}{%
\begin{tikzpicture}[
  >=Stealth,
  font=\sffamily,
  node distance=0.5cm,
  box/.style={
    rounded corners=5pt, draw=#1, line width=1pt,
    fill=#1!8, minimum width=3.8cm, minimum height=0.6cm,
    align=center, font=\sffamily\footnotesize
  },
  modebox/.style={
    rounded corners=4pt, draw=#1, line width=1pt,
    fill=#1!10, minimum width=1.1cm, minimum height=0.5cm,
    align=center, font=\sffamily\scriptsize\bfseries
  },
  arr/.style={->, line width=0.8pt, color=gray!55},
  tag/.style={font=\sffamily\tiny, color=gray!60, fill=white, inner sep=1pt},
]

\definecolor{cH}{HTML}{0D9488}
\definecolor{cA}{HTML}{7C3AED}
\definecolor{cB}{HTML}{2563EB}
\definecolor{cE}{HTML}{DC2626}
\definecolor{cG}{HTML}{059669}
\definecolor{cW}{HTML}{D97706}

\node[box=gray, fill=gray!5, draw=gray!60] (input) at (0,0) {\textbf{User Message}};

\node[box=cH, below=of input] (feat) {Feature Extraction};
\node[tag, right=0pt of feat.east, anchor=west] {1 Flash call};

\node[box=cW, below=of feat] (s1) {\textbf{Stage 1:} Embedding + Soft Penalties};
\node[tag, right=0pt of s1.east, anchor=west] {cosine sim $\geq$ 0.45};

\node[box=cA, below=of s1] (s2) {\textbf{Stage 2:} Questionnaire Disambiguation};
\node[tag, right=0pt of s2.east, anchor=west] {40\%\,emb + 60\%\,Q\&A};

\node[box=cG, below=of s2] (exec) {Agent Execute + Style Inject};
\node[tag, right=0pt of exec.east, anchor=west] {task-adaptive};

\node[box=cE, below=of exec] (guard) {Hallucination Guard};
\node[tag, right=0pt of guard.east, anchor=west] {3-layer};

\node[modebox=cG, below left=0.5cm and -0.1cm of guard] (auto) {auto};
\node[modebox=cW, below=0.5cm of guard] (suggest) {suggest};
\node[modebox=gray, below right=0.5cm and -0.1cm of guard] (none) {none};

\node[font=\sffamily\tiny, color=cG, below=0pt of auto] {deliver};
\node[font=\sffamily\tiny, color=cW, below=0pt of suggest] {offer};
\node[font=\sffamily\tiny, color=gray, below=0pt of none] {standard};

\draw[arr] (input) -- (feat);
\draw[arr] (feat) -- (s1);
\draw[arr] (s1) -- (s2);
\draw[arr] (s2) -- (exec);
\draw[arr] (exec) -- (guard);

\draw[arr] (guard.south) -- ++(0,-0.15) -| (auto.north);
\draw[arr] (guard.south) -- (suggest.north);
\draw[arr] (guard.south) -- ++(0,-0.15) -| (none.north);

\begin{scope}[on background layer]
  \node[rounded corners=8pt, draw=cB, line width=1.2pt, dashed,
        fill=cB!3, fit=(feat)(s1)(s2)(exec)(guard),
        inner xsep=8pt, inner ysep=5pt,
        label={[font=\sffamily\scriptsize\bfseries, color=cB]above left:Active Memory Layer}] {};
\end{scope}

\end{tikzpicture}
}%

%% file: evaluation.tex

\resizebox{\linewidth}{!}{%
\begin{tikzpicture}[
  >=Stealth,
  font=\sffamily,
  node distance=0.45cm,
  agent/.style={
    rounded corners=5pt, draw=#1, line width=1pt,
    fill=#1!8, minimum width=2.2cm, minimum height=0.55cm,
    align=center, font=\sffamily\footnotesize\bfseries
  },
  turnbox/.style={
    rounded corners=3pt, fill=#1!12, draw=#1!40, line width=0.5pt,
    minimum width=0.55cm, minimum height=0.22cm, inner sep=2pt,
  },
  arr/.style={->, line width=0.8pt, color=gray!55},
  tag/.style={font=\sffamily\tiny, color=gray!60},
]

\definecolor{cB}{HTML}{2563EB}
\definecolor{cE}{HTML}{DC2626}
\definecolor{cG}{HTML}{059669}
\definecolor{cW}{HTML}{D97706}
\definecolor{cGray}{HTML}{6B7280}

\node[rounded corners=5pt, draw=gray!60, line width=1pt, fill=gray!5,
      minimum width=3.2cm, minimum height=0.55cm,
      font=\sffamily\footnotesize\bfseries] (open) at (0,0)
      {Same Opening Message};

\coordinate (fork) at (0,-0.5);
\coordinate (leftfork) at (-1.8,-0.5);
\coordinate (rightfork) at (1.8,-0.5);

\draw[arr, -] (open.south) -- (fork);
\draw[arr, -] (leftfork) -- (rightfork);
\draw[arr] (leftfork) -- ++(0,-0.3);
\draw[arr] (rightfork) -- ++(0,-0.3);

\node[agent=cGray] (bl) at (-1.8,-1.25) {Baseline};
\node[tag, below=0pt of bl] {Flash, no tools};

\node[turnbox=cW] (t1) at (-1.8,-2.1) {};
\node[turnbox=cW] (t2) at (-1.8,-2.45) {};
\node[turnbox=cW] (t3) at (-1.8,-2.8) {};
\node[turnbox=cW, opacity=0.4] (t4) at (-1.8,-3.15) {};

\draw[gray!30, line width=0.5pt] (bl.south) -- ++(0,-0.2) -- (t1.north);
\draw[gray!30, line width=0.5pt] (t1.south) -- (t2.north);
\draw[gray!30, line width=0.5pt] (t2.south) -- (t3.north);
\draw[gray!30, line width=0.5pt] (t3.south) -- (t4.north);

\node[font=\sffamily\scriptsize\bfseries, color=cW, below=3pt of t4] (bllabel) {3--4 turns};

\node[agent=cB] (aug) at (1.8,-1.25) {Augmented};
\node[tag, below=0pt of aug] {Flash + swarm};

\node[rounded corners=2pt, draw=cB!40, fill=cB!8,
      font=\sffamily\tiny\color{cB}, inner sep=2pt] (swarm) at (1.8,-2.1) {swarm activates};

\node[turnbox=cG] (at1) at (1.8,-2.55) {};

\node[font=\sffamily\scriptsize\bfseries, color=cG] (goal) at (1.8,-2.95) {\checkmark\ GOAL\_REACHED};

\draw[gray!30, line width=0.5pt] (aug.south) -- ++(0,-0.2) -- (swarm.north);
\draw[gray!30, line width=0.5pt] (swarm.south) -- (at1.north);
\draw[gray!30, line width=0.5pt] (at1.south) -- (goal.north);

\node[font=\sffamily\scriptsize\bfseries, color=cG, below=3pt of goal] (auglabel) {1--2 turns};

\coordinate (jfork) at (0,-4.0);
\draw[arr, -] (bllabel.south) -- ++(0,-0.2) -| (jfork);
\draw[arr, -] (auglabel.south) -- ++(0,-0.15) -| (jfork);
\draw[arr] (jfork) -- ++(0,-0.25);

\node[rounded corners=6pt, draw=cE, line width=1.2pt, fill=cE!6,
      minimum width=4cm, minimum height=0.9cm, align=center,
      font=\sffamily\footnotesize] (judge) at (0,-4.8)
      {\textbf{LLM Judge} --- Gemini 3.1 Pro\\[1pt]
       {\scriptsize 3 dims $\times$ 1--4 scale $\cdot$ declares winner}};

\node[rounded corners=2pt, draw=cE!40, fill=white,
      font=\sffamily\tiny\bfseries\color{cE}, inner sep=2pt]
      at (-1.3,-5.55) {Accuracy};
\node[rounded corners=2pt, draw=cE!40, fill=white,
      font=\sffamily\tiny\bfseries\color{cE}, inner sep=2pt]
      at (0,-5.55) {Helpfulness};
\node[rounded corners=2pt, draw=cE!40, fill=white,
      font=\sffamily\tiny\bfseries\color{cE}, inner sep=2pt]
      at (1.4,-5.55) {Personalization};

\end{tikzpicture}
}%

%% file: main.bib
@article{Packer2024,
  author    = {Charles Packer and Sarah Wooders and Kevin Lin and Vivian Fang and Shishir G. Patil and Ion Stoica and Joseph E. Gonzalez},
  title     = {{MemGPT}: Towards {LLMs} as Operating Systems},
  journal   = {CoRR},
  volume    = {abs/2310.08560},
  year      = {2023},
  note      = {arXiv:2310.08560}
}

@inproceedings{Lewis2020,
  author    = {Patrick S. H. Lewis and Ethan Perez and Aleksandra Piktus and Fabio Petroni and Vladimir Karpukhin and Naman Goyal and Heinrich K{\"{u}}ttler and Mike Lewis and Wen{-}tau Yih and Tim Rockt{\"{a}}schel and Sebastian Riedel and Douwe Kiela},
  title     = {Retrieval-Augmented Generation for Knowledge-Intensive {NLP} Tasks},
  booktitle = {Advances in Neural Information Processing Systems 33 (NeurIPS)},
  year      = {2020}
}

@inproceedings{Zhang2018,
  author    = {Saizheng Zhang and Emily Dinan and Jack Urbanek and Arthur Szlam and Douwe Kiela and Jason Weston},
  title     = {Personalizing Dialogue Agents: {I} have a dog, do you have pets too?},
  booktitle = {Proceedings of the 56th Annual Meeting of the Association for Computational Linguistics (ACL)},
  year      = {2018},
  pages     = {2204--2213}
}

@inproceedings{Salemi2024,
  author    = {Alireza Salemi and Sheshera Mysore and Michael Bendersky and Hamed Zamani},
  title     = {{LaMP}: When Large Language Models Meet Personalization},
  booktitle = {Proceedings of the 62nd Annual Meeting of the Association for Computational Linguistics (ACL)},
  year      = {2024}
}

@inproceedings{Wang2023longmem,
  author    = {Weizhi Wang and Li Dong and Hao Cheng and Xiaodong Liu and Xifeng Yan and Jianfeng Gao and Furu Wei},
  title     = {Augmenting Language Models with Long-Term Memory},
  booktitle = {Advances in Neural Information Processing Systems 36 (NeurIPS)},
  year      = {2023}
}

@inproceedings{Mysore2023,
  author    = {Sheshera Mysore and Zhuoran Lu and Mengting Wan and Longqi Yang and Bahareh Sarrafzadeh and Steve Menezes and Tina Baghaee and Emmanuel Barajas Gonzalez and Jennifer Neville and Tara Safavi},
  title     = {{PEARL:} Personalizing Large Language Model Writing Assistants with Generation-Calibrated Retrievers},
  booktitle = {Proceedings of the 1st Workshop on Customizable NLP (CustomNLP4U)},
  pages     = {198--219},
  year      = {2024}
}

@inproceedings{Shinn2023,
  author    = {Noah Shinn and Federico Cassano and Ashwin Gopinath and Karthik Narasimhan and Shunyu Yao},
  title     = {Reflexion: Language Agents with Verbal Reinforcement Learning},
  booktitle = {Advances in Neural Information Processing Systems 36 (NeurIPS)},
  year      = {2023}
}

@article{Zhang2024survey,
  author    = {Zeyu Zhang and Quanyu Dai and Xiaohe Bo and Chen Ma and Rui Li and Xu Chen and Jieming Zhu and Zhenhua Dong and Ji{-}Rong Wen},
  title     = {A Survey on the Memory Mechanism of Large Language Model-based Agents},
  journal   = {ACM Transactions on Information Systems},
  volume    = {43},
  number    = {6},
  year      = {2025},
  doi       = {10.1145/3748302}
}

@inproceedings{AMem2025,
  author    = {Wujiang Xu and Zujie Liang and Kai Mei and Hang Gao and Juntao Tan and Yongfeng Zhang},
  title     = {{A-Mem}: Agentic Memory for {LLM} Agents},
  booktitle = {Advances in Neural Information Processing Systems (NeurIPS)},
  year      = {2025}
}

@article{Wu2025procedural,
  author    = {Yaxiong Wu and Yongyue Zhang},
  title     = {Agent Skills from the Perspective of Procedural Memory: A Survey},
  journal   = {TechRxiv},
  year      = {2025},
  doi       = {10.36227/techrxiv.176857932.25697838}
}

@inproceedings{Zhou2024lats,
  author    = {Andy Zhou and Kai Yan and Michal Shlapentokh{-}Rothman and Haohan Wang and Yu{-}Xiong Wang},
  title     = {Language Agent Tree Search Unifies Reasoning, Acting, and Planning in Language Models},
  booktitle = {Proceedings of the 41st International Conference on Machine Learning (ICML)},
  year      = {2024}
}

@article{Wang2023voyager,
  author    = {Guanzhi Wang and Yuqi Xie and Yunfan Jiang and Ajay Mandlekar and Chaowei Xiao and Yuke Zhu and Linxi Fan and Anima Anandkumar},
  title     = {Voyager: An Open-Ended Embodied Agent with Large Language Models},
  journal   = {Transactions on Machine Learning Research (TMLR)},
  year      = {2024}
}

@article{SoKSkills2026,
  author    = {Yanna Jiang and Delong Li and Haiyu Deng and Baihe Ma and Xu Wang and Qin Wang and Guangsheng Yu},
  title     = {{SoK:} Agentic Skills --- Beyond Tool Use in {LLM} Agents},
  journal   = {CoRR},
  volume    = {abs/2602.20867},
  year      = {2026}
}

@inproceedings{Yang2024sweagent,
  author    = {John Yang and Carlos E. Jimenez and Alexander Wettig and Kilian Lieret and Shunyu Yao and Karthik Narasimhan and Ofir Press},
  title     = {{SWE}-agent: Agent-Computer Interfaces Enable Automated Software Engineering},
  booktitle = {Advances in Neural Information Processing Systems 37 (NeurIPS)},
  year      = {2024}
}

@article{Wu2023,
  author    = {Qingyun Wu and Gagan Bansal and Jieyu Zhang and Yiran Wu and Beibin Li and Erkang Zhu and Li Jiang and Xiaoyun Zhang and Shaokun Zhang and Jiale Liu and Ahmed Hassan Awadallah and Ryen W. White and Doug Burger and Chi Wang},
  title     = {AutoGen: Enabling Next-Gen {LLM} Applications via Multi-Agent Conversation},
  journal   = {CoRR},
  volume    = {abs/2308.08155},
  year      = {2023}
}

@inproceedings{Hong2024,
  author    = {Sirui Hong and Mingchen Zhuge and Jonathan Chen and Xiawu Zheng and Yuheng Cheng and Ceyao Zhang and Jinlin Wang and Zili Wang and Steven Ka Shing Yau and Zijuan Lin and Liyang Zhou and Chenyu Ran and Lingfeng Xiao and Chenglin Wu and J{\"{u}}rgen Schmidhuber},
  title     = {{MetaGPT}: Meta Programming for {A} Multi-Agent Collaborative Framework},
  booktitle = {Proceedings of the 12th International Conference on Learning Representations (ICLR)},
  year      = {2024}
}

@inproceedings{Li2023camel,
  author    = {Guohao Li and Hasan Abed Al Kader Hammoud and Hani Itani and Dmitrii Khizbullin and Bernard Ghanem},
  title     = {{CAMEL:} Communicative Agents for ``Mind'' Exploration of Large Language Model Society},
  booktitle = {Advances in Neural Information Processing Systems 36 (NeurIPS)},
  year      = {2023}
}

@inproceedings{Chen2023agentverse,
  author    = {Weize Chen and Yusheng Su and Jingwei Zuo and Cheng Yang and Chenfei Yuan and Chi-Min Chan and Heyang Yu and Yaxi Lu and Yi-Hsin Hung and Chen Qian and Yujia Qin and Xin Cong and Ruobing Xie and Zhiyuan Liu and Maosong Sun and Jie Zhou},
  title     = {{AgentVerse}: Facilitating Multi-Agent Collaboration and Exploring Emergent Behaviors},
  booktitle = {Proceedings of the 12th International Conference on Learning Representations (ICLR)},
  year      = {2024}
}

@article{Li2024moreagents,
  author    = {Junyou Li and Qin Zhang and Yangbin Yu and Qiang Fu and Deheng Ye},
  title     = {More Agents Is All You Need},
  journal   = {Transactions on Machine Learning Research (TMLR)},
  year      = {2024},
  note      = {arXiv:2402.05120}
}

@inproceedings{Zheng2023,
  author    = {Lianmin Zheng and Wei{-}Lin Chiang and Ying Sheng and Siyuan Zhuang and Zhanghao Wu and Yonghao Zhuang and Zi Lin and Zhuohan Li and Dacheng Li and Eric P. Xing and Hao Zhang and Joseph E. Gonzalez and Ion Stoica},
  title     = {Judging {LLM}-as-a-Judge with {MT-Bench} and {Chatbot Arena}},
  booktitle = {Advances in Neural Information Processing Systems 36 (NeurIPS)},
  year      = {2023}
}

@article{Dubois2024,
  author    = {Yann Dubois and Bal{\'{a}}zs Galambosi and Percy Liang and Tatsunori B. Hashimoto},
  title     = {Length-Controlled {AlpacaEval:} {A} Simple Way to Debias Automatic Evaluators},
  journal   = {CoRR},
  volume    = {abs/2404.04475},
  year      = {2024}
}

@inproceedings{Wang2024faireval,
  author    = {Peiyi Wang and Lei Li and Liang Chen and Zefan Cai and Dawei Zhu and Binghuai Lin and Yunbo Cao and Lingpeng Kong and Qi Liu and Tianyu Liu and Zhifang Sui},
  title     = {Large Language Models are not Fair Evaluators},
  booktitle = {Proceedings of the 62nd Annual Meeting of the Association for Computational Linguistics (ACL)},
  year      = {2024}
}

@inproceedings{Zhao2025personalens,
  author    = {Zheng Zhao and Clara Vania and Subhradeep Kayal and Naila Khan and Shay B. Cohen and Emine Yilmaz},
  title     = {{PersonaLens}: A Benchmark for Personalization Evaluation in Conversational {AI} Assistants},
  booktitle = {Findings of the Association for Computational Linguistics (ACL)},
  year      = {2025},
  note      = {arXiv:2506.09902}
}

@article{Chen2026skvm,
  author    = {Le Chen and Erhu Feng and Yubin Xia and Haibo Chen},
  title     = {{SkVM}: Revisiting Language {VM} for Skills across Heterogeneous {LLMs} and Harnesses},
  journal   = {CoRR},
  volume    = {abs/2604.03088},
  year      = {2026}
}

@inproceedings{dou2025simulatorarena,
  title={SimulatorArena: Are User Simulators Reliable Proxies for Multi-Turn Evaluation of AI Assistants?},
  author={Dou, Yao and Galley, Michel and Peng, Baolin and Kedzie, Chris and Cai, Weixin and Ritter, Alan and Quirk, Chris and Xu, Wei and Gao, Jianfeng},
  booktitle={Proceedings of the 2025 Conference on Empirical Methods in Natural Language Processing},
  pages={35200--35278},
  year={2025}
}
